\documentclass[aps,prc,twocolumn,superscriptaddress,epsfig]{revtex4}
\usepackage{graphicx}
\begin{document}

%\preprint{}

\title{Gamma-ray spectroscopy of $^{38}_{17}$Cl using grazing reactions}


\author{D.~O'Donnell}
%\email{david.o'donnell@stfc.ac.uk}
\altaffiliation[Present address:]{ STFC Daresbury Laboratory, Warrington, WA4 4AD, United Kingdom}
\affiliation{School of Engineering and Science, University of the West of Scotland, Paisley, PA1 2BE, United
Kingdom}

\author{R.~Chapman}
\affiliation{School of Engineering and Science, University of the West of Scotland, Paisley, PA1 2BE, United
Kingdom}

\author{X.~Liang}
\affiliation{School of Engineering and Science, University of the West of Scotland, Paisley, PA1 2BE, United
Kingdom}

\author{F.~Azaiez}
\affiliation{IPN, IN2P3-CNRS and Universit\'{e} Paris-Sud, F-91406 Orsay Cedex, France}

\author{F.~Haas}
\affiliation{IPHC, CNRS-IN2P3 and Universit\'{e} Louis Pasteur, F-67037 Strasbourg Cedex 2, France}

\author{S.~Beghini}
\affiliation{INFN Sezione di Padova, Universit\`{a} di Padova, I35131 Padova, Italy}

\author{B.~R.~Behera}
\affiliation{INFN, Laboratori Nazionali di Legnaro, I-35020 Legnaro, Padova, Italy}

\author{M.~Burns}
\affiliation{School of Engineering and Science, University of the West of Scotland, Paisley, PA1 2BE, United
Kingdom}

\author{E.~Caurier}
\affiliation{IPHC, CNRS-IN2P3 and Universit\'{e} Louis Pasteur, F-67037 Strasbourg Cedex 2, France}

\author{L.~Corradi}
\affiliation{INFN, Laboratori Nazionali di Legnaro, I-35020 Legnaro, Padova, Italy}

\author{D.~Curien}
\affiliation{IPHC, CNRS-IN2P3 and Universit\'{e} Louis Pasteur, F-67037 Strasbourg Cedex 2, France}

\author{A.~N.~Deacon}
\affiliation{Schuster Laboratory, University of Manchester, Manchester, M13 9PL, United Kingdom}

\author{Z.~S.~Dombr\'{a}di}
\affiliation{ATOMKI, P.O. Box 51, H-4001 Debrecen, Hungary}

\author{E.~Farnea}
\affiliation{INFN Sezione di Padova, Universit\`{a} di Padova, I35131 Padova, Italy}

\author{E.~Fioretto}
\affiliation{INFN, Laboratori Nazionali di Legnaro, I-35020 Legnaro, Padova, Italy}

\author{A.~Hodsdon}
\affiliation{School of Engineering and Science, University of the West of Scotland, Paisley, PA1 2BE, United
Kingdom}

\author{F.~Ibrahim}
\affiliation{IPN, IN2P3-CNRS and Universit\'{e} Paris-Sud, F-91406 Orsay Cedex, France}

\author{A.~Jungclaus}
\affiliation{Dep. de F\'{\i}sica Te\'orica, Universidad Aut\'{o}noma de Madrid, E-28049 Madrid, Spain}

\author{K.~Keyes}
\affiliation{School of Engineering and Science, University of the West of Scotland, Paisley, PA1 2BE, United
Kingdom}

\author{A.~Latina}
\affiliation{INFN, Laboratori Nazionali di Legnaro, I-35020 Legnaro, Padova, Italy}

\author{N.~M\u{a}rginean}
\affiliation{INFN, Laboratori Nazionali di Legnaro, I-35020 Legnaro, Padova, Italy}

\author{G.~Montagnoli}
\affiliation{Dipartimento di Fisica and INFN-Sezione di Padova, Universit\`{a} di Padova, I35131
Padova, Italy}

\author{D.~R.~Napoli}
\affiliation{INFN, Laboratori Nazionali di Legnaro, I-35020 Legnaro, Padova, Italy}

\author{F.~Nowacki}
\affiliation{IPHC, CNRS-IN2P3 and Universit\'{e} Louis Pasteur, F-67037 Strasbourg Cedex 2, France}

\author{J.~Ollier}
\altaffiliation[Present address:]{ STFC Daresbury Laboratory, Warrington, WA4 4AD, United Kingdom}
\affiliation{School of Engineering and Science, University of the West of Scotland, Paisley, PA1 2BE, United
Kingdom}

\author{A.~Papenberg}
\affiliation{School of Engineering and Science, University of the West of Scotland, Paisley, PA1 2BE, United
Kingdom}

\author{G.~Pollarolo}
\affiliation{Dipartimento di Fisica Teorica, Universit\'a di Torino, and Istituto Nazionale di Fisica
Nucleare, Sezione di Torino, I-10125 Torino, Italy}

\author{M.~-D.~Salsac}
\affiliation{IPHC, CNRS-IN2P3 and Universit\'{e} Louis Pasteur, F-67037 Strasbourg Cedex 2, France}

\author{F.~Scarlassara}
\affiliation{Dipartimento di Fisica and INFN-Sezione di Padova, Universit\`{a} di Padova, I35131
Padova, Italy}

\author{J.~F.~Smith}
\affiliation{School of Engineering and Science, University of the West of Scotland, Paisley, PA1 2BE, United
Kingdom}

\author{K.~M.~Spohr}
\affiliation{School of Engineering and Science, University of the West of Scotland, Paisley, PA1 2BE, United
Kingdom}

\author{M.~Stanoiu}
\affiliation{IPN, IN2P3-CNRS and Universit\'{e} Paris-Sud, F-91406 Orsay Cedex, France}

\author{A.~M.~Stefanini}
\affiliation{INFN, Laboratori Nazionali di Legnaro, I-35020 Legnaro, Padova, Italy}

\author{S.~Szilner}
\affiliation{Ruder Bo\v{z}kovi\'{c} Institute, Zagreb, Croatia}

\author{M.~Trotta}
\affiliation{INFN, Laboratori Nazionali di Legnaro, I-35020 Legnaro, Padova, Italy}

\author{D.~Verney}
\affiliation{IPN, IN2P3-CNRS and Universit\'{e} Paris-Sud, F-91406 Orsay Cedex, France}

\author{Z.~M.~Wang}
\affiliation{School of Engineering and Science, University of the West of Scotland, Paisley, PA1 2BE, United
Kingdom}


\date{\today}

\begin{abstract}
Excited states of $^{38}_{17}$Cl$_{21}$ were populated in grazing reactions during the
interaction of a beam of $^{36}_{16}$S$_{20}$ ions of energy 215~MeV with a $^{208}_{82}$Pb$_{126}$
target. The combination of the PRISMA magnetic spectrometer and the CLARA $\gamma$-ray detector array
was used to identify the reaction fragments and to detect their decay via $\gamma$-ray emission. A
level scheme for $^{38}$Cl is presented with tentative spin and parity assignments. The level scheme is
discussed within the context of the systematics of neighboring nuclei and is compared with the results
of state-of-the-art shell model calculations.
\end{abstract}

\pacs{}
%\keywords{}

\maketitle



\section{Introduction}
Nuclei near the $N=20$ shell closure have been the subject of extensive research over the last ten
years. Much discussion and effort has focussed on the neutron-rich species in this region since
the underlying shell structure has been shown to deviate from that of the $\beta$-stable neighboring
nuclei. The isotope $^{38}$Cl with $Z=17$ and $N=21$ has, in its ground state, one neutron
outside the $N=20$ major shell closure and an unpaired proton occupying the 1$d_{3/2}$ orbital.
Thus, $^{38}$Cl, although only one neutron from stability, is an interesting nucleus in which to study cross-shell shell model interactions.

\begin{table}[!b]
\caption{A list of the previously reported experimental studies of $^{38}$Cl.}
\begin{center}
\begin{tabular}[b]{lcc}
\hline
\hline\\
Authors (year) & Ref. & Reaction\\\\
\hline
Paris, Buechner, and Endt (1955) & \cite{Paris} & $^{37}$Cl(d,p)\\
Hoogenboom, Kashy and Buechner (1962) & \cite{Hoogenboom} & $^{37}$Cl(d,p)\\
Rapaport and Buechner (1966) & \cite{Rapaport} & $^{37}$Cl(d,p)\\
Hardy {\it{et al.}} (1970) & \cite{Hardy} & $^{40}$Ar(p,$^3$He)\\
Engelbertink and Olness (1972) & \cite{Engelbertink} & $^{37}$Cl(d,p$\gamma$)\\
Spits and Akkermans (1973) & \cite{Spits} & $^{37}$Cl(n,$\gamma$)\\
Warburton {\it{et al.}} (1986) & \cite{Warburton1} & $^{38}$S $\beta$-decay\\
\hline
\hline
\label{previous_work}
\end{tabular}
\end{center}
\end{table}

The isotope $^{38}$Cl has been the subject of several experimental studies; see Table~\ref{previous_work}.
Although states have been observed up to an excitation energy of 8~MeV, all have
been assigned spins $J\leq5$. Recently, attempts have been made to study high-spin states in the
neutron-rich chlorine isotopes using deep-inelastic reactions with thick targets~\cite{Liang,Liang1,Ollier,Ollier1}. Although these reactions were successful in populating medium-to-high spin-states in neutron-rich chlorine isotopes, such as $^{41}$Cl~\cite{Liang1,Ollier1},
no states were observed to be populated in $^{38}$Cl. This non-observation may be attributed to the short half lives
($<$1ps)~\cite{Spits} of some of the lowest-lying states in $^{38}$Cl.

Several shell-model calculations for $^{38}$Cl have been reported in the literature. Initially, Woods~\cite{Woods} calculated the excitation energies of negative-parity states in $^{38}$Cl up to an
excitation energy of 2681~keV. In that work, two different interactions were used, both of which gave
reasonable agreement with the available experimental data and reproduced the observed ordering of the
states. More recently, a study by Retamosa {\it{et al.}}~\cite{Retamosa} was able to reproduce the
multiplet of states based on the ground-state configuration, $\pi d_{3/2} \otimes \nu f_{7/2}$.

In order to identify higher-spin states ($J>5$) in $^{38}$Cl, the yrast and near-yrast decay
sequences have been studied using grazing reactions with a thin target. The use of a
large solid-angle magnetic spectrometer, in combination with a high-efficiency $\gamma$-ray detector
array, has enabled $\gamma$ rays to be observed in coincidence with reaction products. This
technique has permitted observation of the decay of states which had eluded detection in earlier
deep-inelastic work~\cite{Liang,Liang1,Ollier,Ollier1}. Indeed, with the unambiguous observation of
transitions corresponding to the decay of excited states of $^{38}$Cl in the present work, the data from previous
deep-inelastic studies has been revisited.

\section{Experimental Details}
Excited states of nuclei in the neutron-rich $Z=12-20$ region were populated during the bombardment
of a $^{208}$Pb target by $^{36}$S ions at a laboratory energy of 215~MeV. The beam was delivered by the
combination of the XTU-Tandem and ALPI accelerators at the INFN Legnaro National Laboratory and data
were taken for six days with an average $^{36}$S beam current of $\sim$7~pnA. The target consisted of
isotopically enriched (99.7\%) $^{208}$Pb of thickness 300~$\mu$g/cm$^2$ evaporated onto a carbon
backing which was 20~$\mu$g/cm$^2$ in thickness. Deexcitation $\gamma$ rays from the decay of the
reaction products were detected using {\sc clara} \cite{Gadea}, an array of up to 25 escape-suppressed Ge
clover detectors arranged in a hemispherical configuration and covering the azimuthal angles between
98$^\circ$ and 180$^\circ$, with respect to the entrance of the magnetic spectrometer (see below). For this
particular experiment, 22 of the 25 clover detectors were available.

Projectile-like binary reaction products were separated and identified using the {\sc prisma} magnetic spectrometer~\cite{Stefanini}
which consists of an entrance detector, two magnetic elements and a focal-plane
detector system. The entrance detector is based on a position-sensitive micro-channel plate (25 cm from the
target position) which provides the ($x,y$) coordinates and an initial timing signal for ions entering
the spectrometer~\cite{Montagnoli}. The ions then pass through a quadrupole magnet and are dispersed in
a magnetic dipole before reaching the focal plane. At the focal plane, the ion trajectory and arrival
time are measured as the ion passes through a gas-filled multi-wire parallel plate avalanche counter
before the ion is stopped in a 10 x 4 element ionization chamber~\cite{Beghini}. The ionization chamber
provides measurements of the energy and energy loss of the ion, allowing the atomic number, $Z$, of
the ions to be determined. The time-of-flight and position information are used to trace the ion path
through {\sc prisma} and a correlation between $\gamma$ rays detected in {\sc clara} and ions at the focal plane of
{\sc prisma} is established through coincidence timing measurements. From a knowledge of the velocity vector
of the emitting nucleus, $\gamma$-ray Doppler corrections can be made on an event-by-event basis, with
the energy resolution of $\gamma$-ray photopeaks being typically 0.6\% following correction. The
magnetic spectrometer can be rotated in the reaction plane within a wide angular range of $20^\circ$ to
$120^\circ$, with respect to the beam direction and, during this study, it was positioned at an angle of
56$^\circ$. With a solid angle acceptance of $\approx$80~msr, {\sc prisma} covered a range of angles between
50$^\circ$ and 62$^\circ$, including the grazing angle, $60^\circ$, for this reaction. These features make
{\sc prisma} an ideal tool for studying multi-nucleon transfer reactions, where the differential
cross-sections of reaction products peak at angles close to the grazing angle.

\section{Results}
\begin{figure}
\includegraphics[width=5.5cm,angle=270,clip]{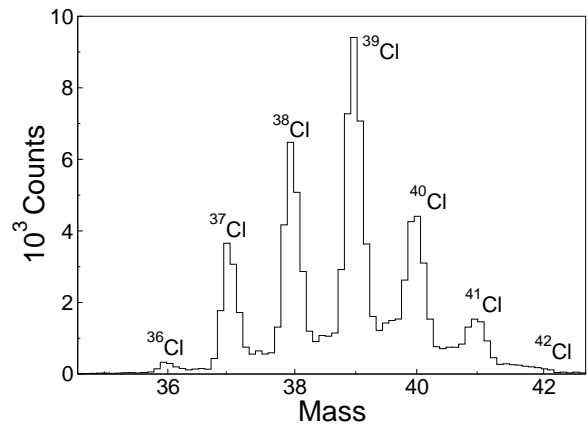}
\caption{Projection of a $\gamma$-{\it A} matrix onto the mass axis showing the isotopes of Cl which were
identified in the present work.}
\label{gamma-mass}
\end{figure}

Figure~\ref{gamma-mass} shows the $\gamma$-{\it A} matrix (projected onto the mass axis) which resulted
from the correlation of $\gamma$ rays and detected Cl ions. Chlorine isotopes with masses in the range
{\it A}=36-42 were populated, with $^{39}$Cl dominating. A software gate can be placed on any mass peak to
obtain a $\gamma$-ray spectrum in coincidence with the reaction product being investigated.

\begin{figure}
\includegraphics[width=6.5cm,angle=270,clip]{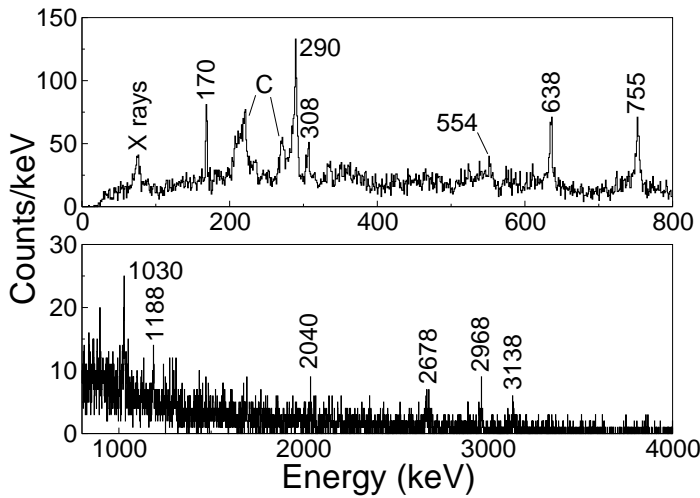}
\caption{Energy spectrum of $\gamma$-rays corresponding to the decay of states of $^{38}$Cl observed in the
present work. The peaks marked with C are contamination resulting from the decay of target-like
reaction fragments associated with $^{38}$Cl.}
\label{gamma-spec}
\end{figure}

The $\gamma$-ray energy spectrum of Figure~\ref{gamma-spec} shows transitions observed in coincidence with the
detection of $^{38}$Cl ions. Unfortunately, this selection procedure does not
ensure that all $\gamma$ rays observed are associated only with the projectile-like reaction product;
$\gamma$-ray photopeaks corresponding to the deexcitation of the unobserved associated target-like products are
also present. This is discussed in more detail in the next section. Measured energies of
the observed $\gamma$ rays and their efficiency-corrected relative intensities are listed in Table
\ref{gamma-rays}. Some peaks were not particularly well defined and the large uncertainty associated with a
number of intensity measurements reflects this. From an inspection of Figure~\ref{gamma-spec}, it is
evident that the statistics are relatively poor meaning that a $\gamma\gamma$ coincidence
analysis was not possible.

\begin{figure}
\includegraphics[width=8cm,clip]{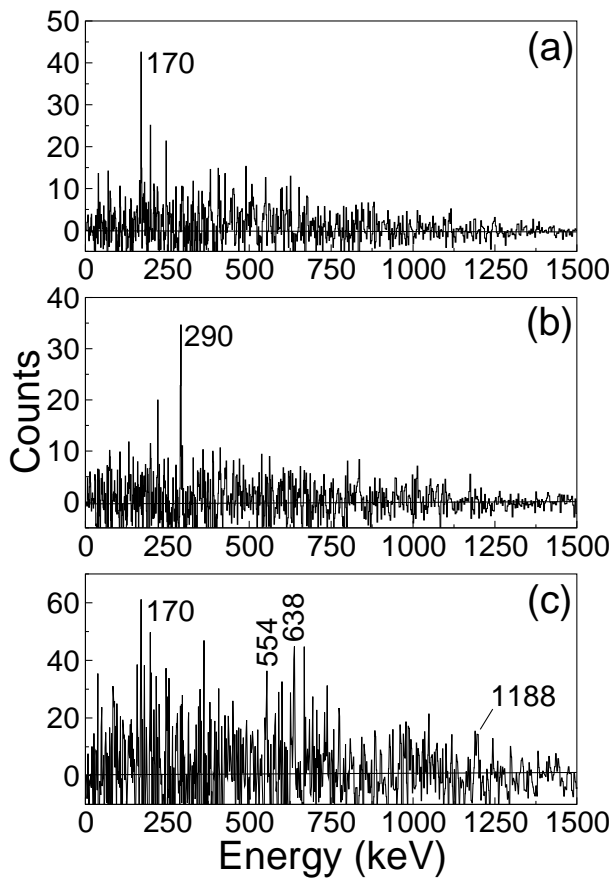}
\caption{Gamma-ray coincidence spectra with gates on (a) 290 and 2678~keV, (b) 170 and 2678~keV and (c) 290 and 2040~keV transitions. The spectra were obtained as a result of a re-analysis of a previous deep-inelastic experiment. See Ref.\cite{Liang,Liang1} and text for more details.}
\label{gamma-gamma_spectra}
\end{figure}

However, the ordering of the observed transitions can be determined by revisiting the data obtained in a
previous deep-inelastic study. In particular, the reaction under discussion involved the bombardment of
a thick target of $^{160}$Gd by 234~MeV $^{37}$Cl ions~\cite{Liang,Liang1}. The deexcitation of the
resulting nuclei by emission of $\gamma$ rays was detected using the {\sc euroball} array~\cite{Simpson} and
$\gamma\gamma\gamma$ correlations were obtained. The population of the $^{38}$Cl channel in this
reaction is expected to be strong, corresponding to the transfer of one neutron from the $^{160}$Gd
target to the $^{37}$Cl projectile. However, as outlined in the introduction, the half lives of the
low-lying excited states of $^{38}$Cl ($5^-:715 $~ms, $3^-:220 $~fs and $4^-:370$~fs~\cite{Endt4}) make
it a particularly difficult nucleus to study using deep-inelastic reactions performed with a thick target since the
decays of the first $J^\pi=3^-$ and $4^-$ states occur in-flight. The resulting Doppler broadening
means that the observed intensity of the transitions from these states is low, with only those
$\gamma$ rays emitted from the small fraction of stopped $^{38}$Cl ions being resolved. Double gated
$\gamma$-ray spectra involving those transitions first identified in the present work are shown in
Figure~\ref{gamma-gamma_spectra}. The coincidence relationships between the transitions of 170, 290 and 2678~keV can be clearly seen from Figures~\ref{gamma-gamma_spectra}(a) and ~\ref{gamma-gamma_spectra}(b) while Figure~\ref{gamma-gamma_spectra}(c) suggests the 290 and 2040~keV transitions are in coincidence with the previously reported transitions of energies 554 and 638~keV.

\begin{table}
\caption{A list of the $\gamma$-rays observed in this work, ordered by excitation energy of decaying
state.}
\begin{center}
\begin{tabular}{ccccc}
\hline
\hline\\
E$_i$ (keV) & E$_{\gamma}$ (keV) & $J^{\pi}_i$ & $J^{\pi}_f$ & I$_{\gamma}$/I$_{755} (\%)$\\\\
\hline
755 & 754.6(3) & $3^-$ & $2^-$ & 100(20)\\
1309 & 637.7(5) & $4^-$ & $5^-$ & 65(13)\\
1309 & 554.3(6) & $4^-$ & $3^-$ & 18(7)\\
1617 & 307.6(5) & $3^-$ & $4^-$ & 18(6)\\
1617 & 862.4(7) & $3^-$ & $3^-$ & 11(2)\\
1785 & 1029.9(5) & $(2,3,4)$ & $3^-$ & 35(6)\\
3349 & 2677.7(7) & $(7^+)$ & $5^-$ & 48(9)\\
3349 & 2039.8(3) & $(7^+)$ & $4^-$ & 15(5)\\
3639 & 290.2(2) & $(5,6)$ & $(7^+)$ & 30(10)\\
3639 & 2968.1(5) & $(5,6)$ & $5^-$ & 37(8)\\
3809 & 169.6(2) & $(4,5,6)$ & $(5,6)$ & 24(5)\\
3809 & 3138.4(6) & $(4,5,6)$ & $5^-$ & 34(8)\\
4827 & 1187.9(6) & $(J \geq 5)$ & $(5,6)$ & 19(5)\\
\hline
\hline
\label{gamma-rays}
\end{tabular}
\end{center}
\end{table}

\begin{figure}
\includegraphics[width=8.75cm,angle=270,clip]{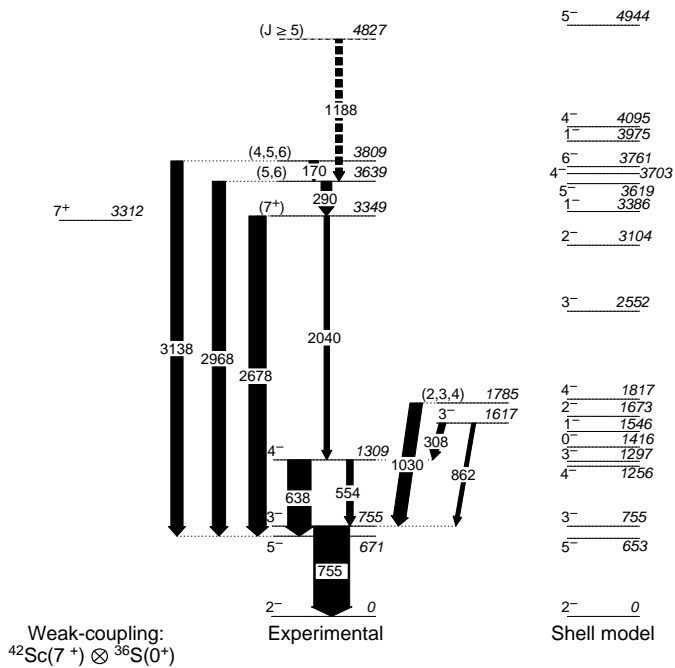}
\caption{The level scheme of $^{38}$Cl based on the present work. Relative $\gamma$-ray intensities are
indicated by the width of the arrows. The results of shell model calculations are presented in addition to weak-coupling calculations. For
states of excitation energy up to 1785 keV, $J^\pi$ values are taken from earlier published works
\cite{Hoogenboom,Engelbertink,Spits}. For states added as a result of the present work, proposed
$J^\pi$ assignments are given. See text for details.}
\label{level_scheme}
\end{figure}

\section{Discussion}
The level scheme for the yrast and near-yrast decay sequence of $^{38}$Cl based on the present work is
compared with the results of recent state-of-the-art shell-model calculations in Figure~\ref{level_scheme}.
The calculations were performed using the shell-model code ANTOINE~\cite{Caurier2}
and involved an inert $^{16}$O core and an effective interaction which is based on results
showing a reduction in the spin-orbit splitting of $f$ and $p$ orbitals in the vicinity of $^{47}$Ar~\cite{Gaudefroy}.
In this particular calculation the valence protons are restricted to occupy the $sd$ shell
and neutron or proton particle-hole excitations across the $N=20$ shell gap are not considered. As a result,
in this $0~\hbar\omega$ valence space, only negative parity states of $^{38}$Cl may be calculated.

States observed in the present work up to an excitation energy of 1785~keV were previously identified
in Refs.~\cite{Engelbertink,Spits}. The 671~keV transition, from the isomeric (715~ms) first excited
$J^\pi=5^-$ state to the ground state, was not observed here since the lifetime of this state is longer than
the flight time of the $^{38}$Cl ions through {\sc prisma}. In this work, the level scheme has been
extended by the addition of four states at energies of 3349, 3639, 3809 and 4997~keV. Seven previously
unobserved transitions with energies of 170, 290, 1188, 2040, 2678, 2968 and 3138 keV have been added
to the level scheme.

The first three excited states of $^{38}$Cl were initially identified by Paris, Buechner and Endt~\cite{Paris}
using the $^{37}$Cl$(d,p)$ reaction. Hoogenboom {\it et al.}~\cite{Hoogenboom} later
studied this reaction in more detail and measured the absolute differential cross sections and angular
distributions of protons. This allowed spins to be inferred and parities to be determined for the
previously observed energy levels, based on the measured orbital angular momentum quantum number of the
transferred neutron and on shell model arguments. The $J^\pi$ assignments, which are given in Figure~\ref{level_scheme},
suggest that the states are part of a quadruplet of levels, which includes the
$^{38}$Cl ground state, based on the shell model configuration $\pi(1d_{\frac{3}{2}})^1\nu(1f_{\frac{7}{2}})^1$.
This is supported by a consideration of states in
$^{40}_{19}$K$_{21}$~\cite{Goldstein}. While $^{38}_{17}$Cl$_{21}$ can be considered as a
particle-particle nucleus, $^{40}_{19}$K$_{21}$ has a particle-hole configuration with two additional
protons occupying the $1d_{\frac{3}{2}}$ orbital. The so-called Pandya transformation~\cite{Pandya} can
be used to calculate the energies of the $\pi(1d_{\frac{3}{2}})^1\nu(1f_{\frac{7}{2}})^1$ multiplet of
$^{38}$Cl states from the spectrum of $\pi(1d_{\frac{3}{2}})^{-1}\nu(1f_{\frac{7}{2}})^1$ states of
$^{40}$K. The success of this method, first reported by Goldstein and Talmi~\cite{Goldstein}, was taken
as proof that the low-lying multiplet in $^{38}$Cl was based on the
$\pi(1d_{\frac{3}{2}})^1\nu(1f_{\frac{7}{2}})^1$ configuration. Subsequent measurements~\cite{Wedberg}
of the magnetic-dipole transitions between the low-lying states of $^{38}$Cl, however, showed some
discrepancies with this simplified picture. It was suggested~\cite{Wedberg} that, in order to explain
the observations adequately, either surprisingly large $(\sim30\%)$ admixtures of
$\pi(2s_{\frac{1}{2}})\nu(1f_{\frac{7}{2}})$ and $\pi(1d_{\frac{3}{2}})\nu(2p_{\frac{3}{2}})$
components are present, or contributions from configurations higher in the $fp$-shell should be
considered. The shell model calculations performed in the present work support this hypothesis. The
first excited $J^\pi=3^-$ state is well reproduced by a state whose wavefunction consists of
$\pi(1d_{\frac{3}{2}})\nu(1f_{\frac{7}{2}})$ ($\sim$55\%), $\pi(1d_{\frac{3}{2}})\nu(2p_{\frac{3}{2}})$
($\sim$21\%) and $\pi(2s_{\frac{1}{2}})\nu(1f_{\frac{7}{2}})$ ($\sim$14\%) components. The other states
of this multiplet are well reproduced by shell model states predominantly ($\geq$67\%) corresponding to
the $\pi(1d_{\frac{3}{2}})\nu(1f_{\frac{7}{2}})$ configuration and having small admixtures of other
configurations, each contributing $\leq 5\%$ of the wavefunction. This may explain the observed
branching ratio in the decay of the 1309~keV $J^\pi=4^-$ state in which the transition to the $J^\pi=5^-$ state is
significantly favored over that to the $3^-$ state, despite both transitions being $M1/E2$ in nature.

\begin{figure}
\includegraphics[angle=270,width=8cm,clip]{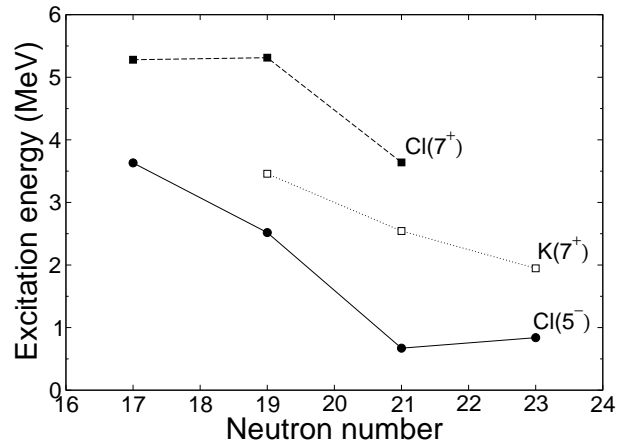}
\caption{Systematic trends of stretched $J^\pi = 5^-$ and $7^+$ states in odd-$A$ Cl and K nuclei.}
\label{systematics}
\end{figure}

In Figure~\ref{systematics}, the $J^\pi = 5^-$ and $7^+$ states of $^{34,36,38}$Cl and $^{38,40}$K are compared.
One notable feature is the dramatic lowering of the $J^\pi=5^-$ state in the chlorine isotopes as the
number of neutrons is increased. The lowering of this state from an energy of 2518~keV in $^{36}$Cl to
671~keV in $^{38}$Cl is what would be expected across a shell closure. As $^{38}$Cl has a single
neutron occupying the $1f_{\frac{7}{2}}$ orbital in its ground state, considerably less energy is
required to produce a $5^-$ state than is necessary in the lower mass isotopes where the neutron has to
be promoted across the $N=20$ shell gap.

Above the $\pi(1d_{\frac{3}{2}})^1\nu(1f_{\frac{7}{2}})^1$ multiplet in $^{38}$Cl, the next two excited states
observed in the present study are the previously established~\cite{Paris,Hoogenboom,Engelbertink,Spits,Warburton1}
$J^\pi=3^-$ and (2,3,4) states at 1617 and 1785~keV, respectively, shown in Figure~\ref{level_scheme}. Relative
intensity measurements of the two $\gamma$-ray transitions depopulating the state at 1617~keV are in good agreement,
within experimental uncertainty, with the previously reported branching ratios~\cite{Engelbertink,Spits}.
Engelbertink and Olness~\cite{Engelbertink} observed a 946~keV transition from the 1617~keV state to
the isomeric $J^\pi=5^-$ state with a branching ratio of $\approx$ 6\% relative to that of the 308~keV
transition. This branching ratio implies that this decay is too weak to be observed in the present study.
In Ref.\cite{Spits}, the state observed at 1785~keV was assigned a tentative $J$ value of 2, 3 or 4 and positive
parity. However, Warburton {\it et al.}~\cite{Warburton1} assigned a tentative $J^\pi=4^-$ value to this state. The
latter assignment seems to be supported by the shell model calculations of this work with
$2^-$ and $4^-$ states calculated to exist in the vicinity of the experimentally observed energy of
1785~keV. Since only negative parity states have been calculated, this is not conclusive. Here, a
$\gamma$-ray of energy 1030~keV is observed decaying from this state to the $J^\pi=3^-$ 755~keV
state. Four additional transitions depopulating the 1785~keV state were first reported by Engelbertink
and Olness~\cite{Engelbertink}. The branching ratios reported in the previous work suggest these transitions are too
weak to be observed in the present work.

Four previously unreported excited states, with excitation energies of 3349, 3639, 3809 and 4827~keV, have
been identified in this work. It has not been possible to determine the multipolarities of the $\gamma$ rays
emitted by these states and, hence, spin and parity could not be established. Nuclei in the vicinity of
$^{38}$Cl can be turned to in order to understand the origins of the observed states. The $J^\pi = 5^-$ and
$7^+$ states in $^{34,36}$Cl have been successfully populated in the past using the $(\alpha, d)$
reaction~\cite{DelVecchio,Nann}. In this reaction the neutron and proton are preferentially transferred
in a state of maximum alignment of angular momentum. The $J^\pi=5^-$ and $7^+$ states in both $^{34}$Cl
and $^{36}$Cl were identified by Nann {\it et al.}~\cite{Nann} as stretched states ($J = \ell_\pi + \ell_\nu + 2s$)
corresponding to the configurations $\pi(1d_{\frac{3}{2}})\nu(1f_{\frac{7}{2}})$ and
$\pi(1f_{\frac{7}{2}})\nu(1f_{\frac{7}{2}})$, respectively. The former configuration corresponds to the
671~keV state of $^{38}_{17}$Cl$_{21}$ while one would expect the latter to be observed as a highly
excited state of this nucleus. As $^{38}$Cl has a neutron occupying the $1f_{\frac{7}{2}}$ orbital in
its ground state, the $\pi(1f_{\frac{7}{2}})\nu(1f_{\frac{7}{2}})$ excited state is expected at a lower
energy than is observed in the lower-mass chlorine isotopes. Indeed, one would expect the evolution of
the $7^+$ states in the even-$A$ chlorine nuclei to behave similarly to that of the $5^-$ states
discussed above. The reduction in energy of the $7^+$ states in K nuclei as the $N=20$ shell gap is crossed
is evident from Fig.~\ref{systematics}. A reduction of 1847~keV is observed in the energies of the $J^\pi=5^-$ states in
moving from $^{36}$Cl to $^{38}$Cl. Assuming a similar decrease in the energy of the $7^+$ state, the
three states observed in this work between 3 and 4~MeV are good candidates. Therefore, it is suggested
that one of the states with energies 3349, 3639 and 3809 keV corresponds to the
$\pi(1f_{\frac{7}{2}})\nu(1f_{\frac{7}{2}})$ configuration and has $J^\pi=7^+$.

A closer inspection of the $\gamma$-ray decay of the states at 3349, 3639 and 3809~keV places
restrictions on the possible spin and parity values. Two transitions were identified in the decay of
the 3349~keV level. This state was observed to decay to the isomeric $J^\pi=5^-$ and to the $J^\pi=4^-$
levels with 76\% and 24\% of the total observed decay probability, respectively. This suggests that the
3349~keV state has $J \ge 3$. However, one would expect this state to have a spin value higher than 3
as the nature of grazing reactions means that yrast states are predominantly populated.
The shell-model calculations show a $J^\pi=1^-$ state at an excitation energy of 3386~keV. However,
according to the arguments above, the observation of a $J^\pi=1^-$ state at such a high excitation
energy in the present work seems unlikely. In addition, the decay from such a state to the $4^-$ and
$5^-$ states would be prohibitively hindered having multipolarities of $M3$ and $E4$, respectively. In
accordance with the above arguments, and those to follow, the 3349~keV state is tentatively assigned a
spin and parity $7^+$.

The 3639~keV state is observed to decay via two $\gamma$ rays with energies 290 and 2968~keV. The
intensity of the 290~keV decay could not be determined accurately as a result of contamination in the
$\gamma$-ray spectrum. This contamination has been identified as the result of $\gamma$-ray transitions
within the complementary fragments $^{206,203}$Tl. The contamination by target-like fragments
contributes to the large background observed in the spectrum of Fig.~\ref{gamma-spec}. The
efficiency-corrected intensities of Table~\ref{gamma-rays} suggest the two transitions are of
comparable strength with the 2968 and 290~keV transitions accounting for 55\% and 45\%, respectively,
of the total observed decay strength from this level. That the two transitions are comparable in strength is
rather unexpected when one considers the $\gamma$-ray energies involved. This could be explained if the
2968~keV decay corresponds to a change in parity while the 290~keV $\gamma$-ray links states of the same
parity. However, the shell-model calculations presented in Fig.~\ref{level_scheme} show that this state
is well reproduced by a $J^\pi=5^-$ state based predominantly (75\%) on a
$\pi[(1d_{\frac{5}{2}})^6(2s_{\frac{1}{2}})^{-1}(1d_{\frac{3}{2}})^2]_{{\frac{5}{2}}^+}\otimes\nu[(1f_{
\frac{7}{2}})^1]_{{\frac{7}{2}}^-}$ configuration. This assignment seems unlikely in light of the, albeit
tentative, $7^+$ assigned to the 3349~keV state. Nonetheless, due to a lack of conclusive evidence and the
conflicting arguments above, the parity of this state remains undetermined but the observed decay
suggests this state most likely corresponds to $J = 5$ or $6$.

The state observed at 3809~keV also decays via two $\gamma$-ray branches; the 170~keV transition (41\%
of the observed decay from this state) to the 3639-keV state and the 3138~keV transition (59\%) to the
671~keV state. A comparison with the shell-model calculations suggests that this state may have spin
and parity of $6^-$ with the dominant (88\%) wavefunction component corresponding to the
$\pi[(1d_{\frac{5}{2}})^6(2s_{\frac{1}{2}})^{-1}(1d_{\frac{3}{2}})^2]_{{\frac{5}{2}}^+}\otimes\nu[(1f_{
\frac{7}{2}})^1]_{{\frac{7}{2}}^-}$ configuration. Once more, the parity of this state can not be
determined but consideration of the assignments to the lower energy states

\begin{figure}
\includegraphics[width=6cm,angle=270,clip]{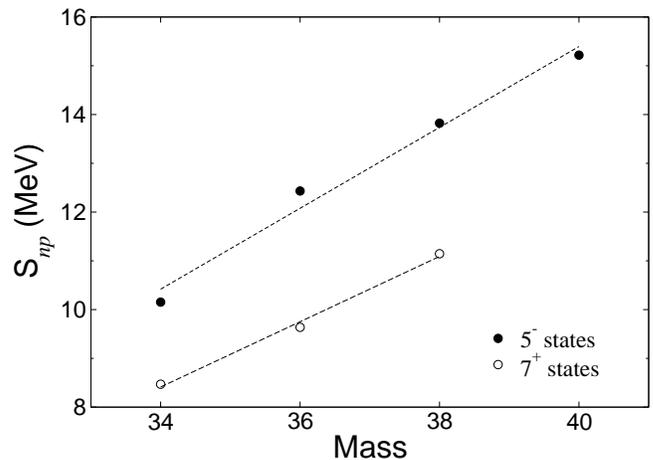}
\caption{Plot of neutron-proton separation energy as a function of mass number for available
$J^\pi=5^-$ and $7^+$ states in even-A chlorine isotopes. The dashed lines represent linear least-squares fits to the data.}
\label{Snp_systematics}
\end{figure}

In the $(\alpha, d)$ experiments~\cite{DelVecchio,Nann} used to populate states with stretched
configurations in $^{34}$Cl and $^{36}$Cl, the neutron-proton $np$ pair, transferred from the
projectile to the target, can be treated as being coupled to $^{32}$S and
$^{34}$S ground state cores, respectively. In this way the separation energy of the $np$ pair,
S$_{np}$, can then be calculated. Fig.~\ref{Snp_systematics} shows the evolution of S$_{np}$ with mass
number for the two stretched states with $J^\pi=5^-$ and $7^+$ for the even-$A$ chlorine isotopes
being discussed here. Inspection of this figure shows a strong linear dependence for both $5^-$ and $7^+$
states. This behavior is commonly interpreted within the Bansal-French-Zamick weak-coupling model~\cite{Bansal,Zamick}
and is consistent with previous analyzes of high-spin states based on two nucleon stretched
configurations~\cite{Rivet,Chan,Ollier2}. A quantitative estimate for the energy at which a $J^\pi = 7^+$ state is expected
in $^{38}$Cl can also be obtained by considering the Bansal-French-Zamick weak-coupling model. Excitations of nuclei relative to an inert
core nucleus can be treated as representing the particle-hole configuration being investigated. This approach has been shown to be successful in reproducing particle-hole excitations in a number of $p$, $sd$ and $fp$ shell nuclei~\cite{Bansal,Sherr,DelVecchio,Bernstein,Manley}. In the present work, the first excited $7^+$ state of $^{38}$Cl was treated as a 2 particle-4 hole configuration with a maximally-aligned neutron-proton $f_{7/2}$ pair outside of a $^{40}_{20}$Ca$_{20}$ core. The particle configuration is manifested in the low-lying $7^+$ excitation of $^{42}_{21}$Sc$_{21}$ while the four proton holes in the $1d_{3/2}$ orbital have been represented by the ground state of $^{36}_{16}$S$_{20}$. Using the formalism and parameters outlined in Ref.~\cite{Sherr}, the energy of the weakly-coupled $7^+$ state in $^{38}$Cl was calculated to be 3312~keV. This result lends further support to the tentative $J^\pi = 7^+$ assigned to the observed state at 3349~keV.

The decay of a $J^\pi=7^+$ state to the $5^-$ level at 671~keV would correspond to a mixed M2/E3 transition while the 2040-keV transition to the 1309-keV $4^-$ state would be of E3 character. In the study of the high-spin states of $^{40}$K, Eggenhuisen {\it et al.}~\cite{Eggenhuisen} observed similar competition between $M2/E3$ ($7^+\rightarrow{5^-}$) and $E3$ ($7^+\rightarrow{4^-}$) transitions indicating that the observed competition in the present work is reasonable.

The observed competition between the low energy transitions of the 3639 and 3809~keV states and their high energy decay to the 671~keV state may be explained if the three states between 3 and 4~MeV are treated as members of a multiplet based on the configuration $\pi(1f_{\frac{7}{2}})\nu(1f_{\frac{7}{2}})$. Since such a configuration would result in states of positive parity, it is suggested that the lowest-lying member of this multiplet would be the odd-{\it J} state with the largest spatial overlap, namely the $J^\pi=7^+$ state. The other two proposed members of the multiplet are most likely to have $J^\pi=4^+,5^+$ or $6^+$ since one state decays to the $7^+$ state and both states decay to the 671-keV $5^-$ state. In their study of nucleon-nucleon matrix elements, Schiffer and True~\cite{Schiffer} showed that the $7^+$ state is indeed the lowest state observed in a multiplet based on the $(1f_{\frac{7}{2}})^2$ configuration; the conclusions of Ref.\cite{Schiffer} are supported by a more recent study performed by Daehnick~\cite{Daehnick}. Of the $J^\pi$ assignments considered in the present discussion, the $5^+$ state is the next most lowered with the $6^+$ member of the multiplet experiencing the least lowering of all members. Based on this $(1f_{\frac{7}{2}})^2$ multiplet argument it is suggested that the 3639~keV state has $J^\pi=5^+$ while the state observed at 3809~keV may have $J^\pi=4^+$ or $6^+$. Such spin and parity assignments would mean that the high energy transitions of energy 2968 and 3138~keV would be $E1$ in nature while the 170 and 290~keV `intra-multiplet' transitions would be $M1/E2$ and $E2$ transitions, respectively. The non-observation of a transition between the 3809 and 3349~keV states suggests that the spin and parity of the 3809~keV state is more likely $4^+$ and not $6^+$, assuming the above argument is valid.

The state of highest excitation energy observed in the present work is the tentative
4827~keV state. Its position in the level scheme can be supported by the observation that its decay is
observed in coincidence with the 290~ and 2040~keV transitions (Figure~\ref{gamma-gamma_spectra}(c) but
is not observed in coincidence with the 170~keV $\gamma$-ray (Figures~\ref{gamma-gamma_spectra}(b). The shell-model calculations reproduce this state reasonably well with a third excited $5^-$ state
consisting of a large ($\sim$63\%)
$\pi[(2s_{\frac{1}{2}})^{1}(1d_{\frac{3}{2}})^{2}]\nu(1f_{\frac{7}{2}})$ component; however,
significant contributions from the
$\pi[(2s_{\frac{1}{2}})^{0}(1d_{\frac{3}{2}})^{3}]\nu(1f_{\frac{7}{2}})$ $(\sim14\%)$ and
$\pi[(1d_{\frac{5}{2}})^{-1}(2s_{\frac{1}{2}})^{1}(1d_{\frac{3}{2}})^{3}]\nu(1f_{\frac{7}{2}})$
$(\sim11\%)$ components are also necessary in order to explain this state. However, if the three states
immediately below this state belong to the proposed $(1f_{\frac{7}{2}})^2$ multiplet and the states at
3639~keV has $J^\pi=5^+$ then the decay via the 1188~keV transition suggests that the 4827~keV state
has $J\geq 5$.

\section{Summary}
Grazing reactions, involving a $^{36}$S beam at an energy of 215~MeV on a $^{208}$Pb target, have been successfully used to populate excited states of $^{38}$Cl. The CLARA $\gamma$-ray detector array in combination with the magnetic spectrometer PRISMA were used to identify the isotope and study the decay of its excited states via the emission of $\gamma$ rays. In total, thirteen $\gamma$ rays were identified leading to the identification of nine excited states up to an excitation energy of $\approx$ 5~MeV. The previously reported yrast level scheme was extended by considering the $\gamma$-ray energies and intensities. An analysis of $\gamma\gamma\gamma$ coincidence data from a previous thick-target deep-inelastic experiment in which $^{38}$Cl was populated, has supported the level scheme presented here. Shell-model calculations corresponding to an improved $sdfp$ interaction have been performed for $^{38}$Cl. The spins and parities of the newly identified states have been tentatively assigned, based on systematics of the neighboring isotopes and guided by the results of our shell-model calculations. A candidate for the stretched $J^\pi=7^+$ configuration $\pi(1f_{\frac{7}{2}})\nu(1f_{\frac{7}{2}})$ has been identified based on the systematics of np separation energies of $7^+$ states populated in the $(\alpha,d)$ reaction and weak-coupling calculations. Evidence for the observation of other members of the multiplet resulting from the proposed $\pi(1f_{\frac{7}{2}})\nu(1f_{\frac{7}{2}})$ configuration is also discussed.

\begin{acknowledgments}
This work was supported in part by the EPSRC (UK) and by the European Union under contract number
RII3-CT-2004-506065. Six of us (DO'D, MB, AD, AH, KK and AP) acknowledge financial support from the
EPSRC. AJ acknowledges financial support from the Spanish Ministerio de Educaci\'on y Ciencia with the
programa Ram\'on y Cajal and project FPA2005-00696. The contribution of the accelerator and
target-fabrication staff at the INFN Legnaro National Laboratory is gratefully acknowledged.
\end{acknowledgments}

\bibliography{bibliography}

\begin{thebibliography}{37}
\expandafter\ifx\csname natexlab\endcsname\relax\def\natexlab#1{#1}\fi
\expandafter\ifx\csname bibnamefont\endcsname\relax
  \def\bibnamefont#1{#1}\fi
\expandafter\ifx\csname bibfnamefont\endcsname\relax
  \def\bibfnamefont#1{#1}\fi
\expandafter\ifx\csname citenamefont\endcsname\relax
  \def\citenamefont#1{#1}\fi
\expandafter\ifx\csname url\endcsname\relax
  \def\url#1{\texttt{#1}}\fi
\expandafter\ifx\csname urlprefix\endcsname\relax\def\urlprefix{URL }\fi
\providecommand{\bibinfo}[2]{#2}
\providecommand{\eprint}[2][]{\url{#2}}

\bibitem[{\citenamefont{Paris et~al.}(1955)\citenamefont{Paris, Buechner, and
  Endt}}]{Paris}
\bibinfo{author}{\bibfnamefont{C.~H.} \bibnamefont{Paris}},
  \bibinfo{author}{\bibfnamefont{W.~W.} \bibnamefont{Buechner}},
  \bibnamefont{and} \bibinfo{author}{\bibfnamefont{P.~M.} \bibnamefont{Endt}},
  \bibinfo{journal}{Phys. Rev.} \textbf{\bibinfo{volume}{100}},
  \bibinfo{pages}{1317} (\bibinfo{year}{1955}).

\bibitem[{\citenamefont{Hoogenboom et~al.}(1962)\citenamefont{Hoogenboom,
  Kashy, and Buechner}}]{Hoogenboom}
\bibinfo{author}{\bibfnamefont{A.~M.} \bibnamefont{Hoogenboom}},
  \bibinfo{author}{\bibfnamefont{E.}~\bibnamefont{Kashy}}, \bibnamefont{and}
  \bibinfo{author}{\bibfnamefont{W.~W.} \bibnamefont{Buechner}},
  \bibinfo{journal}{Phys. Rev.} \textbf{\bibinfo{volume}{128}},
  \bibinfo{pages}{305} (\bibinfo{year}{1962}).

\bibitem[{\citenamefont{Rapaport and Buechner}(1966)}]{Rapaport}
\bibinfo{author}{\bibfnamefont{J.}~\bibnamefont{Rapaport}} \bibnamefont{and}
  \bibinfo{author}{\bibfnamefont{W.~W.} \bibnamefont{Buechner}},
  \bibinfo{journal}{Nucl. Phys.} \textbf{\bibinfo{volume}{83}},
  \bibinfo{pages}{80} (\bibinfo{year}{1966}).

\bibitem[{\citenamefont{Hardy et~al.}(1970)\citenamefont{Hardy, Brunnader, and
  Cerny}}]{Hardy}
\bibinfo{author}{\bibfnamefont{J.~C.} \bibnamefont{Hardy}},
  \bibinfo{author}{\bibfnamefont{H.}~\bibnamefont{Brunnader}},
  \bibnamefont{and} \bibinfo{author}{\bibfnamefont{J.}~\bibnamefont{Cerny}},
  \bibinfo{journal}{Phys. Rev. C} \textbf{\bibinfo{volume}{1}},
  \bibinfo{pages}{561} (\bibinfo{year}{1970}).

\bibitem[{\citenamefont{Engelbertink and Olness}(1972)}]{Engelbertink}
\bibinfo{author}{\bibfnamefont{G.~A.~P.} \bibnamefont{Engelbertink}}
  \bibnamefont{and} \bibinfo{author}{\bibfnamefont{J.~W.}
  \bibnamefont{Olness}}, \bibinfo{journal}{Phys. Rev. C}
  \textbf{\bibinfo{volume}{5}}, \bibinfo{pages}{431} (\bibinfo{year}{1972}).

\bibitem[{\citenamefont{Spits and Akkermans}(1973)}]{Spits}
\bibinfo{author}{\bibfnamefont{A.~M.~J.} \bibnamefont{Spits}} \bibnamefont{and}
  \bibinfo{author}{\bibfnamefont{J.~A.} \bibnamefont{Akkermans}},
  \bibinfo{journal}{Nucl. Phys} \textbf{\bibinfo{volume}{A215}},
  \bibinfo{pages}{260} (\bibinfo{year}{1973}).

\bibitem[{\citenamefont{Warburton et~al.}(1986)\citenamefont{Warburton,
  Alburger, Becker, Brown, and Raman}}]{Warburton1}
\bibinfo{author}{\bibfnamefont{E.~K.} \bibnamefont{Warburton}},
  \bibinfo{author}{\bibfnamefont{D.~E.} \bibnamefont{Alburger}},
  \bibinfo{author}{\bibfnamefont{J.~A.} \bibnamefont{Becker}},
  \bibinfo{author}{\bibfnamefont{B.~A.} \bibnamefont{Brown}}, \bibnamefont{and}
  \bibinfo{author}{\bibfnamefont{S.}~\bibnamefont{Raman}},
  \bibinfo{journal}{Phys. Rev. C} \textbf{\bibinfo{volume}{34}},
  \bibinfo{pages}{1031} (\bibinfo{year}{1986}).

\bibitem[{\citenamefont{Liang}(2002)}]{Liang}
\bibinfo{author}{\bibfnamefont{X.}~\bibnamefont{Liang}}, Ph.D. thesis,
  \bibinfo{school}{University of Paisley} (\bibinfo{year}{2002}).

\bibitem[{\citenamefont{Liang et~al.}(2002)\citenamefont{Liang, Chapman, Haas,
  Spohr, Bednarczyk, Campbell, Dagnall, Davison, de~Angelis, Duch\^ene
  et~al.}}]{Liang1}
\bibinfo{author}{\bibfnamefont{X.}~\bibnamefont{Liang}},
  \bibinfo{author}{\bibfnamefont{R.}~\bibnamefont{Chapman}},
  \bibinfo{author}{\bibfnamefont{F.}~\bibnamefont{Haas}},
  \bibinfo{author}{\bibfnamefont{K.~M.} \bibnamefont{Spohr}},
  \bibinfo{author}{\bibfnamefont{P.}~\bibnamefont{Bednarczyk}},
  \bibinfo{author}{\bibfnamefont{S.~M.} \bibnamefont{Campbell}},
  \bibinfo{author}{\bibfnamefont{P.~J.} \bibnamefont{Dagnall}},
  \bibinfo{author}{\bibfnamefont{M.}~\bibnamefont{Davison}},
  \bibinfo{author}{\bibfnamefont{G.}~\bibnamefont{de~Angelis}},
  \bibinfo{author}{\bibfnamefont{G.}~\bibnamefont{Duch\^ene}},
  \bibnamefont{et~al.}, \bibinfo{journal}{Phys. Rev. C}
  \textbf{\bibinfo{volume}{66}}, \bibinfo{pages}{037301 }
  (\bibinfo{year}{2002}).

\bibitem[{\citenamefont{Ollier}(2004)}]{Ollier}
\bibinfo{author}{\bibfnamefont{J.}~\bibnamefont{Ollier}}, Ph.D. thesis,
  \bibinfo{school}{University of Paisley} (\bibinfo{year}{2004}).

\bibitem[{\citenamefont{Ollier et~al.}(2004)\citenamefont{Ollier, Chapman,
  Liang, Labiche, Spohr, Davison, , de~Angelis, Axiotis, Kr{\"{o}}ll
  et~al.}}]{Ollier1}
\bibinfo{author}{\bibfnamefont{J.}~\bibnamefont{Ollier}},
  \bibinfo{author}{\bibfnamefont{R.}~\bibnamefont{Chapman}},
  \bibinfo{author}{\bibfnamefont{X.}~\bibnamefont{Liang}},
  \bibinfo{author}{\bibfnamefont{M.}~\bibnamefont{Labiche}},
  \bibinfo{author}{\bibfnamefont{K.~M.} \bibnamefont{Spohr}},
  \bibinfo{author}{\bibfnamefont{M.}~\bibnamefont{Davison}}, ,
  \bibinfo{author}{\bibfnamefont{G.}~\bibnamefont{de~Angelis}},
  \bibinfo{author}{\bibfnamefont{M.}~\bibnamefont{Axiotis}},
  \bibinfo{author}{\bibfnamefont{T.}~\bibnamefont{Kr{\"{o}}ll}},
  \bibnamefont{et~al.}, \bibinfo{journal}{Eur. Phys. J. A}
  \textbf{\bibinfo{volume}{20}}, \bibinfo{pages}{111} (\bibinfo{year}{2004}).

\bibitem[{\citenamefont{Woods}(1985)}]{Woods}
\bibinfo{author}{\bibfnamefont{C.~L.} \bibnamefont{Woods}},
  \bibinfo{journal}{Nuc. Phys.} \textbf{\bibinfo{volume}{A451}},
  \bibinfo{pages}{413} (\bibinfo{year}{1985}).

\bibitem[{\citenamefont{Retamosa et~al.}(1997)\citenamefont{Retamosa, Caurier,
  Nowacki, and Poves}}]{Retamosa}
\bibinfo{author}{\bibfnamefont{J.}~\bibnamefont{Retamosa}},
  \bibinfo{author}{\bibfnamefont{E.}~\bibnamefont{Caurier}},
  \bibinfo{author}{\bibfnamefont{F.}~\bibnamefont{Nowacki}}, \bibnamefont{and}
  \bibinfo{author}{\bibfnamefont{A.}~\bibnamefont{Poves}},
  \bibinfo{journal}{Phys. Rev. C} \textbf{\bibinfo{volume}{55}},
  \bibinfo{pages}{1266} (\bibinfo{year}{1997}).

\bibitem[{\citenamefont{Gadea et~al.}(2004)\citenamefont{Gadea, Napoli,
  de~Angelis, Menegazzo, Stefanini, Corradi, Axiotis, Berti, Fioretto,
  Kr{\"{o}}ll et~al.}}]{Gadea}
\bibinfo{author}{\bibfnamefont{A.}~\bibnamefont{Gadea}},
  \bibinfo{author}{\bibfnamefont{D.~R.} \bibnamefont{Napoli}},
  \bibinfo{author}{\bibfnamefont{G.}~\bibnamefont{de~Angelis}},
  \bibinfo{author}{\bibfnamefont{R.}~\bibnamefont{Menegazzo}},
  \bibinfo{author}{\bibfnamefont{A.~M.} \bibnamefont{Stefanini}},
  \bibinfo{author}{\bibfnamefont{L.}~\bibnamefont{Corradi}},
  \bibinfo{author}{\bibfnamefont{M.}~\bibnamefont{Axiotis}},
  \bibinfo{author}{\bibfnamefont{L.}~\bibnamefont{Berti}},
  \bibinfo{author}{\bibfnamefont{E.}~\bibnamefont{Fioretto}},
  \bibinfo{author}{\bibfnamefont{T.}~\bibnamefont{Kr{\"{o}}ll}},
  \bibnamefont{et~al.}, \bibinfo{journal}{Eur. Phys. J. A}
  \textbf{\bibinfo{volume}{20}}, \bibinfo{pages}{193} (\bibinfo{year}{2004}).

\bibitem[{\citenamefont{Stefanini et~al.}(2002)\citenamefont{Stefanini,
  Corradi, Maron, Pisent, Trotta, Vinodkumar, Beghini, Montagnoli, Scarlassara,
  Segato et~al.}}]{Stefanini}
\bibinfo{author}{\bibfnamefont{A.~M.} \bibnamefont{Stefanini}},
  \bibinfo{author}{\bibfnamefont{L.}~\bibnamefont{Corradi}},
  \bibinfo{author}{\bibfnamefont{G.}~\bibnamefont{Maron}},
  \bibinfo{author}{\bibfnamefont{A.}~\bibnamefont{Pisent}},
  \bibinfo{author}{\bibfnamefont{M.}~\bibnamefont{Trotta}},
  \bibinfo{author}{\bibfnamefont{A.~M.} \bibnamefont{Vinodkumar}},
  \bibinfo{author}{\bibfnamefont{S.}~\bibnamefont{Beghini}},
  \bibinfo{author}{\bibfnamefont{G.}~\bibnamefont{Montagnoli}},
  \bibinfo{author}{\bibfnamefont{F.}~\bibnamefont{Scarlassara}},
  \bibinfo{author}{\bibfnamefont{G.~F.} \bibnamefont{Segato}},
  \bibnamefont{et~al.}, \bibinfo{journal}{Nucl. Phys.}
  \textbf{\bibinfo{volume}{A701}}, \bibinfo{pages}{217} (\bibinfo{year}{2002}).

\bibitem[{\citenamefont{Montagnoli et~al.}(2005)\citenamefont{Montagnoli,
  Stefanini, Trotta, Beghini, Bettini, Scarlassara, Schiavon, Corradi, Behera,
  Fioretto et~al.}}]{Montagnoli}
\bibinfo{author}{\bibfnamefont{G.}~\bibnamefont{Montagnoli}},
  \bibinfo{author}{\bibfnamefont{A.~M.} \bibnamefont{Stefanini}},
  \bibinfo{author}{\bibfnamefont{M.}~\bibnamefont{Trotta}},
  \bibinfo{author}{\bibfnamefont{S.}~\bibnamefont{Beghini}},
  \bibinfo{author}{\bibfnamefont{M.}~\bibnamefont{Bettini}},
  \bibinfo{author}{\bibfnamefont{F.}~\bibnamefont{Scarlassara}},
  \bibinfo{author}{\bibfnamefont{V.}~\bibnamefont{Schiavon}},
  \bibinfo{author}{\bibfnamefont{L.}~\bibnamefont{Corradi}},
  \bibinfo{author}{\bibfnamefont{B.~R.} \bibnamefont{Behera}},
  \bibinfo{author}{\bibfnamefont{E.}~\bibnamefont{Fioretto}},
  \bibnamefont{et~al.}, \bibinfo{journal}{Nucl. Inst. and Meth in Phys. Res. A}
  \textbf{\bibinfo{volume}{547}}, \bibinfo{pages}{455} (\bibinfo{year}{2005}).

\bibitem[{\citenamefont{Beghini et~al.}(2005)\citenamefont{Beghini, Corradi,
  Fioretto, Gadea, Latina, Montagnoli, Scarlassara, Stefanini, Szilner, Trotta
  et~al.}}]{Beghini}
\bibinfo{author}{\bibfnamefont{S.}~\bibnamefont{Beghini}},
  \bibinfo{author}{\bibfnamefont{L.}~\bibnamefont{Corradi}},
  \bibinfo{author}{\bibfnamefont{E.}~\bibnamefont{Fioretto}},
  \bibinfo{author}{\bibfnamefont{A.}~\bibnamefont{Gadea}},
  \bibinfo{author}{\bibfnamefont{A.}~\bibnamefont{Latina}},
  \bibinfo{author}{\bibfnamefont{G.}~\bibnamefont{Montagnoli}},
  \bibinfo{author}{\bibfnamefont{F.}~\bibnamefont{Scarlassara}},
  \bibinfo{author}{\bibfnamefont{A.~M.} \bibnamefont{Stefanini}},
  \bibinfo{author}{\bibfnamefont{S.}~\bibnamefont{Szilner}},
  \bibinfo{author}{\bibfnamefont{M.}~\bibnamefont{Trotta}},
  \bibnamefont{et~al.}, \bibinfo{journal}{Nucl. Inst. and Meth in Phys. Res. A}
  \textbf{\bibinfo{volume}{551}}, \bibinfo{pages}{364} (\bibinfo{year}{2005}).

\bibitem[{\citenamefont{Simpson}(1997)}]{Simpson}
\bibinfo{author}{\bibfnamefont{J.}~\bibnamefont{Simpson}}, \bibinfo{journal}{Z.
  Phys. A} \textbf{\bibinfo{volume}{358}}, \bibinfo{pages}{139}
  (\bibinfo{year}{1997}).

\bibitem[{\citenamefont{Endt}(1990)}]{Endt4}
\bibinfo{author}{\bibfnamefont{P.~M.} \bibnamefont{Endt}},
  \bibinfo{journal}{Nucl. Phys.} \textbf{\bibinfo{volume}{521}},
  \bibinfo{pages}{1} (\bibinfo{year}{1990}).

\bibitem[{\citenamefont{Caurier et~al.}(2005)\citenamefont{Caurier,
  Mart\'inez-Pinedo, Nowacki, Poves, and Zuker}}]{Caurier2}
\bibinfo{author}{\bibfnamefont{E.}~\bibnamefont{Caurier}},
  \bibinfo{author}{\bibfnamefont{G.}~\bibnamefont{Mart\'inez-Pinedo}},
  \bibinfo{author}{\bibfnamefont{F.}~\bibnamefont{Nowacki}},
  \bibinfo{author}{\bibfnamefont{A.}~\bibnamefont{Poves}}, \bibnamefont{and}
  \bibinfo{author}{\bibfnamefont{A.~P.} \bibnamefont{Zuker}},
  \bibinfo{journal}{Rev. Mod. Phys.} \textbf{\bibinfo{volume}{77}},
  \bibinfo{pages}{427} (\bibinfo{year}{2005}).

\bibitem[{\citenamefont{Gaudefroy et~al.}(2006)\citenamefont{Gaudefroy, Sorlin,
  Beaumel, Blumenfeld, Dombr\'{a}~di, Fortier, Franchoo, G\'{e}~lin, Gibelin,
  Gr\'{e}~vy Hammache et~al.}}]{Gaudefroy}
\bibinfo{author}{\bibfnamefont{L.}~\bibnamefont{Gaudefroy}},
  \bibinfo{author}{\bibfnamefont{O.}~\bibnamefont{Sorlin}},
  \bibinfo{author}{\bibfnamefont{D.}~\bibnamefont{Beaumel}},
  \bibinfo{author}{\bibfnamefont{Y.}~\bibnamefont{Blumenfeld}},
  \bibinfo{author}{\bibfnamefont{Z.}~\bibnamefont{Dombr\'{a}~di}},
  \bibinfo{author}{\bibfnamefont{S.}~\bibnamefont{Fortier}},
  \bibinfo{author}{\bibfnamefont{S.}~\bibnamefont{Franchoo}},
  \bibinfo{author}{\bibfnamefont{M.}~\bibnamefont{G\'{e}~lin}},
  \bibinfo{author}{\bibfnamefont{J.}~\bibnamefont{Gibelin}},
  \bibinfo{author}{\bibfnamefont{F.}~\bibnamefont{Gr\'{e}~vy Hammache}},
  \bibnamefont{et~al.}, \bibinfo{journal}{Phys. Rev. Lett.}
  \textbf{\bibinfo{volume}{97}}, \bibinfo{pages}{092501}
  (\bibinfo{year}{2006}).

\bibitem[{\citenamefont{Goldstein and Talmi}(1956)}]{Goldstein}
\bibinfo{author}{\bibfnamefont{S.}~\bibnamefont{Goldstein}} \bibnamefont{and}
  \bibinfo{author}{\bibfnamefont{I.}~\bibnamefont{Talmi}},
  \bibinfo{journal}{Phys. Rev.} \textbf{\bibinfo{volume}{102}},
  \bibinfo{pages}{589} (\bibinfo{year}{1956}).

\bibitem[{\citenamefont{Pandya}(1956)}]{Pandya}
\bibinfo{author}{\bibfnamefont{S.~P.} \bibnamefont{Pandya}},
  \bibinfo{journal}{Phys. Rev.} \textbf{\bibinfo{volume}{103}},
  \bibinfo{pages}{956} (\bibinfo{year}{1956}).

\bibitem[{\citenamefont{Wedberg et~al.}(1970)\citenamefont{Wedberg, Beard,
  Puttaswamy, and Williams}}]{Wedberg}
\bibinfo{author}{\bibfnamefont{G.~H.} \bibnamefont{Wedberg}},
  \bibinfo{author}{\bibfnamefont{G.~B.} \bibnamefont{Beard}},
  \bibinfo{author}{\bibfnamefont{N.~G.} \bibnamefont{Puttaswamy}},
  \bibnamefont{and} \bibinfo{author}{\bibfnamefont{N.}~\bibnamefont{Williams}},
  \bibinfo{journal}{Phys. Rev. Lett.} \textbf{\bibinfo{volume}{25}},
  \bibinfo{pages}{1352} (\bibinfo{year}{1970}).

\bibitem[{\citenamefont{Del~Vecchio et~al.}(1976)\citenamefont{Del~Vecchio,
  Kouzes, and Sherr}}]{DelVecchio}
\bibinfo{author}{\bibfnamefont{R.~M.} \bibnamefont{Del~Vecchio}},
  \bibinfo{author}{\bibfnamefont{R.~T.} \bibnamefont{Kouzes}},
  \bibnamefont{and} \bibinfo{author}{\bibfnamefont{R.}~\bibnamefont{Sherr}},
  \bibinfo{journal}{Nucl. Phys. A} \textbf{\bibinfo{volume}{265}},
  \bibinfo{pages}{220} (\bibinfo{year}{1976}).

\bibitem[{\citenamefont{Nann et~al.}(1977)\citenamefont{Nann, Chien, Saha, and
  Wildenthal}}]{Nann}
\bibinfo{author}{\bibfnamefont{H.}~\bibnamefont{Nann}},
  \bibinfo{author}{\bibfnamefont{W.~S.} \bibnamefont{Chien}},
  \bibinfo{author}{\bibfnamefont{A.}~\bibnamefont{Saha}}, \bibnamefont{and}
  \bibinfo{author}{\bibfnamefont{B.~H.} \bibnamefont{Wildenthal}},
  \bibinfo{journal}{Phys. Rev. C} \textbf{\bibinfo{volume}{15}},
  \bibinfo{pages}{1959} (\bibinfo{year}{1977}).

\bibitem[{\citenamefont{Bansal and French}(1964)}]{Bansal}
\bibinfo{author}{\bibfnamefont{R.~K.} \bibnamefont{Bansal}} \bibnamefont{and}
  \bibinfo{author}{\bibfnamefont{J.~B.} \bibnamefont{French}},
  \bibinfo{journal}{Phys. Lett.} \textbf{\bibinfo{volume}{11}},
  \bibinfo{pages}{145} (\bibinfo{year}{1964}).

\bibitem[{\citenamefont{Zamick}(1965)}]{Zamick}
\bibinfo{author}{\bibfnamefont{L.}~\bibnamefont{Zamick}},
  \bibinfo{journal}{Phys. Lett.} \textbf{\bibinfo{volume}{19}},
  \bibinfo{pages}{580} (\bibinfo{year}{1965}).

\bibitem[{\citenamefont{Rivet et~al.}(1966)\citenamefont{Rivet, Pehl, Cerny,
  and Harvey}}]{Rivet}
\bibinfo{author}{\bibfnamefont{E.}~\bibnamefont{Rivet}},
  \bibinfo{author}{\bibfnamefont{R.~H.} \bibnamefont{Pehl}},
  \bibinfo{author}{\bibfnamefont{J.}~\bibnamefont{Cerny}}, \bibnamefont{and}
  \bibinfo{author}{\bibfnamefont{B.~G.} \bibnamefont{Harvey}},
  \bibinfo{journal}{Phys. Rev.} \textbf{\bibinfo{volume}{141}},
  \bibinfo{pages}{1021} (\bibinfo{year}{1966}).

\bibitem[{\citenamefont{Chan}(1987)}]{Chan}
\bibinfo{author}{\bibfnamefont{T.~U.} \bibnamefont{Chan}},
  \bibinfo{journal}{Phys. Rev. C} \textbf{\bibinfo{volume}{36}},
  \bibinfo{pages}{838} (\bibinfo{year}{1987}).

\bibitem[{\citenamefont{Ollier et~al.}(2005)\citenamefont{Ollier, Chapman,
  Liang, Labiche, Spohr, Davison, de~Angelis, Axiotis, Kr{\"{o}}ll\aa, Napoli
  et~al.}}]{Ollier2}
\bibinfo{author}{\bibfnamefont{J.}~\bibnamefont{Ollier}},
  \bibinfo{author}{\bibfnamefont{R.}~\bibnamefont{Chapman}},
  \bibinfo{author}{\bibfnamefont{X.}~\bibnamefont{Liang}},
  \bibinfo{author}{\bibfnamefont{M.}~\bibnamefont{Labiche}},
  \bibinfo{author}{\bibfnamefont{K.-M.} \bibnamefont{Spohr}},
  \bibinfo{author}{\bibfnamefont{M.}~\bibnamefont{Davison}},
  \bibinfo{author}{\bibfnamefont{G.}~\bibnamefont{de~Angelis}},
  \bibinfo{author}{\bibfnamefont{M.}~\bibnamefont{Axiotis}},
  \bibinfo{author}{\bibfnamefont{T.}~\bibnamefont{Kr{\"{o}}ll\aa}},
  \bibinfo{author}{\bibfnamefont{D.~R.} \bibnamefont{Napoli}},
  \bibnamefont{et~al.}, \bibinfo{journal}{Phys. Rev. C}
  \textbf{\bibinfo{volume}{71}}, \bibinfo{pages}{034316}
  (\bibinfo{year}{2005}).

\bibitem[{\citenamefont{Sherr et~al.}(1974)\citenamefont{Sherr, Kouzes, and
  Del~Vecchio}}]{Sherr}
\bibinfo{author}{\bibfnamefont{R.}~\bibnamefont{Sherr}},
  \bibinfo{author}{\bibfnamefont{R.~T.} \bibnamefont{Kouzes}},
  \bibnamefont{and}
  \bibinfo{author}{\bibfnamefont{R.}~\bibnamefont{Del~Vecchio}},
  \bibinfo{journal}{Phys. Lett.} \textbf{\bibinfo{volume}{52B}},
  \bibinfo{pages}{401} (\bibinfo{year}{1974}).

\bibitem[{\citenamefont{Bernstein}(1972)}]{Bernstein}
\bibinfo{author}{\bibfnamefont{A.~M.} \bibnamefont{Bernstein}},
  \bibinfo{journal}{Ann. Phys.} \textbf{\bibinfo{volume}{69}},
  \bibinfo{pages}{19} (\bibinfo{year}{1972}).

\bibitem[{\citenamefont{Manley et~al.}(1990)\citenamefont{Manley, Millener,
  Berman, Bertozzi, Buti, Finn, Hersman, Hyde-Wright, Hynes, Kelly
  et~al.}}]{Manley}
\bibinfo{author}{\bibfnamefont{D.~M.} \bibnamefont{Manley}},
  \bibinfo{author}{\bibfnamefont{D.~J.} \bibnamefont{Millener}},
  \bibinfo{author}{\bibfnamefont{B.~L.} \bibnamefont{Berman}},
  \bibinfo{author}{\bibfnamefont{W.}~\bibnamefont{Bertozzi}},
  \bibinfo{author}{\bibfnamefont{T.~N.} \bibnamefont{Buti}},
  \bibinfo{author}{\bibfnamefont{J.~M.} \bibnamefont{Finn}},
  \bibinfo{author}{\bibfnamefont{F.~W.} \bibnamefont{Hersman}},
  \bibinfo{author}{\bibfnamefont{C.~E.} \bibnamefont{Hyde-Wright}},
  \bibinfo{author}{\bibfnamefont{M.~V.} \bibnamefont{Hynes}},
  \bibinfo{author}{\bibfnamefont{J.~J.} \bibnamefont{Kelly}},
  \bibnamefont{et~al.}, \bibinfo{journal}{Phys. Rev. C}
  \textbf{\bibinfo{volume}{41}}, \bibinfo{pages}{448} (\bibinfo{year}{1990}).

\bibitem[{\citenamefont{Eggenhuisen et~al.}(1977)\citenamefont{Eggenhuisen,
  Ekstr{\"{o}}m, Engelbertink, Aarts, and Hermans}}]{Eggenhuisen}
\bibinfo{author}{\bibfnamefont{H.~H.} \bibnamefont{Eggenhuisen}},
  \bibinfo{author}{\bibfnamefont{L.~P.} \bibnamefont{Ekstr{\"{o}}m}},
  \bibinfo{author}{\bibfnamefont{G.~A.~P.} \bibnamefont{Engelbertink}},
  \bibinfo{author}{\bibfnamefont{H.~J.~M.} \bibnamefont{Aarts}},
  \bibnamefont{and} \bibinfo{author}{\bibfnamefont{J.~A.~J.}
  \bibnamefont{Hermans}}, \bibinfo{journal}{Nucl. Phys.}
  \textbf{\bibinfo{volume}{A285}}, \bibinfo{pages}{167} (\bibinfo{year}{1977}).

\bibitem[{\citenamefont{Schiffer and True}(1976)}]{Schiffer}
\bibinfo{author}{\bibfnamefont{J.~P.} \bibnamefont{Schiffer}} \bibnamefont{and}
  \bibinfo{author}{\bibfnamefont{W.~W.} \bibnamefont{True}},
  \bibinfo{journal}{Rev. Mod. Phys.} \textbf{\bibinfo{volume}{48}},
  \bibinfo{pages}{191} (\bibinfo{year}{1976}).

\bibitem[{\citenamefont{Daehnick}(1983)}]{Daehnick}
\bibinfo{author}{\bibfnamefont{W.~W.} \bibnamefont{Daehnick}},
  \bibinfo{journal}{Phys. Rep.} \textbf{\bibinfo{volume}{96}},
  \bibinfo{pages}{317} (\bibinfo{year}{1983}).

\end{thebibliography}

\end{document}